\documentclass[useAMS,usenatbib]{mn2e}
\usepackage{epsfig,graphicx,latexsym,amsmath,amssymb}
\usepackage{natbib}
\usepackage{hyperref}
\usepackage{mathrsfs}
\usepackage{lastpage}

\bibpunct[,]{(}{)}{;}{a}{}{,}

\newcommand {\Msun}{\ensuremath{M^{}_{\odot}}}

\newcommand {\kms}{\ensuremath{\mathrm{km\,s}^{-1}}}

\usepackage{color}

\title[Spin matters: EMRI event rates]
      {The role of the supermassive black hole spin in the estimation of the EMRI event rate}

\author[P. Amaro-Seoane et al] 
{Pau Amaro-Seoane$^{1}$
                        \thanks{E-mail: Pau.Amaro-Seoane@aei.mpg.de (PAS)}, 
Carlos F. Sopuerta$^{2}$ 
\& Marc Dewi Freitag$^{3,\,4}$
   \\
$^{1}$Max Planck Institut f\"ur Gravitationsphysik
(Albert-Einstein-Institut), D-14476 Potsdam, Germany\\
$^{2}$Institut de Ci{\`e}ncies de l'Espai (CSIC-IEEC), Campus UAB,
Torre C5 parells, 08193 Bellaterra, Spain\\
$^{3}$Institute of Astronomy, Madingley Road, Cambridge, CB3 0HA, UK\\
$^{4}$Gymnase de Nyon, Route de Divonne 8, 1260 Nyon, Switzerland
}

\begin{document}

\date{\today}

\pagerange{\pageref{firstpage}--\pageref{lastpage}} \pubyear{2012}

\maketitle

\label{firstpage}

\begin{abstract}
One of the main channels of interactions in galactic nuclei between stars and
the central massive black hole (MBH) is the gradual inspiral of compact
remnants into the MBH due to the emission of gravitational radiation. This
process is known as an ``Extreme Mass Ratio Inspiral'' (EMRI).  Previous works
about the estimation of how many events space observatories such as LISA will
be able to observe during its operational time differ in orders of magnitude,
due to the complexity of the problem. Nevertheless, a common result to all
investigations is that the possibility that a compact object merges with the
MBH after only one intense burst of GWs is much more likely than a slow
adiabatic inspiral, an EMRI.  The later is referred to as a ``plunge'' because
the compact object dives into the MBH, crosses the horizon and is lost as a
probe of strong gravity for eLISA.  The event rates for plunges are orders of
magnitude larger than slow inspirals.  On the other hand, nature MBH's are most
likely Kerr and the magnitude of the spin has been sized up to be high. We
calculate the number of periapsis passages that a compact object set on to an
extremely radial orbit goes through before being actually swallowed by the Kerr
MBH and we then translate it into an event rate for a LISA-like observatory,
such as the proposed ESA mission eLISA/NGO.  We prove that a ``plunging''
compact object is conceptually indistinguishable from an adiabatic, slow
inspiral; plunges spend on average up to hundred of thousands of cycles in the
bandwidth of the detector for a two years mission.  This has an important
impact on the event rate, enhancing in some cases significantly, depending on
the spin of the MBH and the inclination: If the orbit of the EMRI is prograde,
the effective size of the MBH becomes smaller the larger the spin is, whilst if
retrograde, it becomes bigger.  However, this situation is not symmetric,
resulting in an effective enhancement of the rates. The effect of vectorial
resonant relaxation on the sense of the orbit does not affect the enhancement.
Moreover, it has been recently proved that the production of
low-eccentricity EMRIs is severely blocked by the presence of a blockade in the 
rate at which orbital angular momenta change
takes place. This is the result of relativistic precession on to the stellar potential torques
and hence affects EMRIs originating via resonant relaxation at distances of about $\sim 10^{-2}$ pc from the MBH.
Since high-eccentricity EMRIs are a result of two-body relaxation, they are not affected by this phenomenon.
Therefore we predict that eLISA EMRI event rates will be dominated by high-eccentricity binaries, as we present here.
\end{abstract}

\maketitle

\section{Motivation}\label{sec.motivation}

We know, mostly through high-resolution observations of the kinematics of stars
and gas, that most, if not all, nearby bright galaxies harbour a dark, massive,
compact object at their centres. \citep{FF04,Kormendy04}. The most spectacular
case is our own galaxy, the Milky Way.  By tracking and interpreting the
stellar dynamics at the centre of our galaxy, we have the most well-established
evidence for the existence of a massive black hole (MBH)
\citep{EisenhauerEtAl05,GhezEtAl05,GhezEtAl08,GillessenEtAl09}.  Observations
of other galaxies indicate that the masses of MBH can reach a few billion solar
masses ($\Msun$). The existence of such a MBH population in the present-day
universe is strongly supported by So{\l}tan's argument that the average mass
density of these MBHs agrees with expectations from integrated luminosity of
quasars \citep{Soltan82,YT02}.  Many correlations linking the MBH's mass and
overall properties of its host spheroid (bulge or elliptical galaxy) have been
discovered. The tightest are with the spheroid mass \citep{HR04}, its velocity
dispersion ($M-\sigma$ relation, \citealt{TremaineEtAl02}) and degree of
concentration \citep{EGC04}.  Consequently, understanding the origin and
evolution of these MBHs necessitates their study in the context of their
surrounding stellar systems.

The ideal probe of these regions is the gravitational radiation (GWs) that is
emitted by some compact stars very close to the black holes, and which will be
surveyed by eLISA/NGO \citep[evolved Laser Interferometer Space Antenna / New
Gravitational Wave Observatory][]{Amaro-SeoaneEtAl2012,Amaro-SeoaneEtAl2012b}.
This mission will scrutinise the range of masses fundamental to the
understanding of the origin and growth of supermassive black holes; i.e. MBHs
with masses below $10^7\,M_{\odot}$. 

\section{EMRIs and direct plunges}\label{sec.ermiplunges}

For a binary of a MBH and a stellar black hole to be in a LISA-like band, it has
to have a frequency of between roughly $1$ and $10^{-5}$ Hz. The emission of
GWs is more efficient as they approach the LSO, so that the observatory will detect the
sources when they are close to the LSO line. The total mass required to observe
systems with frequencies between $0.1$ Hz and $10^{-4}$ Hz is of $10^4 -
10^7\,M^{}_{\odot}$. For masses larger than $10^7\,M^{}_{\odot}$ the frequencies
close to the LSO will be too low, so that their detection will be very
difficult. On the other hand, for a total mass of less than $10^3\,M^{}_{\odot}$
we could in principal detect them at an early stage, but then the amplitude of
the GW would be rather low.

To interact closely with the central MBH, stars have to find themselves on
``loss-cone'' orbits, which are orbits elongated enough to have a very close-in
periapsis \citep{FR76,LS77,AS01}. The rate of tidal disruptions can be
established semi-/analytically if the phase space distribution of stars around
the MBH is known \citep[][ for estimates in models of observed nearby
nuclei]{MT99,SU99,WM04}. 
To account for the complex influence of mass
segregation, collisions and the evolution of the nucleus over billions of
years, detailed numerical simulations are required, however
\citep{DDC87a,DDC87b,MCD91,FB02b,BME04b,PauTesi04,FAK06a,KhalEtAl07,PretoAmaroSeoane10,Amaro-SeoanePreto11}.

As the star spirals
down towards the MBH, it has many opportunities to be deflected back by
two-body encounters onto a ``safer orbit'' \citep{AH03,Amaro-SeoaneEtAl07},
hence even the definition of a loss cone is not straightforward. Once again,
the problem is compounded by the effects of mass segregation and resonant
relaxation, to mention two main complications. As a result, considerable
uncertainties are attached to the (semi-)analytical predictions of capture
rates and orbital parameters of EMRIs.

Naively one could assume that the inspiral time is dominated by GW emission and
that if this is shorter than a Hubble time, the compact object will become an
EMRI.  This is wrong, because one has to take into account the relaxation of
the stellar system.  Whilst it certainly can increase the eccentricity of the
compact object, it can also perturb the orbit and circularise it, so that the
required time to inspiral in, $t^{}_{\rm GW}$, becomes larger than a Hubble time. 
The condition for the small compact object to be an EMRI is that it is on an orbit 
for which $t^{}_{\rm GW} \ll (1-e)\,t^{}_{\rm r}$ \citep{Amaro-SeoaneEtAl07}, with
$t^{}_{\rm r}$ the {\em local} relaxation time. When the binary has a
semi-major axis for which the condition is not fulfilled, the small compact
object will have to be already on a so-called ``plunging orbit'', with $e\ge
e_{\rm plunge} \equiv 1-4\,R^{}_{\rm Schw}/a$, where $R^{}_{\rm Schw}$ is the
Schwarzschild radius of the MBH, i.e. $R^{}_{\rm Schw} =
2GM^{}_{\bullet}/c^{2}$, with $M^{}_{\bullet}$ being the MBH mass.  It has been
claimed a number of times by different authors that this would result in a too
short burst of gravitational radiation which could only be detected if it was
originated in our own Galactic Center \citep{HopmanFreitagLarson07}, because one needs a
coherent integration of some few thousands repeated passages through the
periapsis in the eLISA bandwidth.

Therefore, such ``plunging'' objects would then be lost for the GW signal,
since they would be plunging ``directly'' through the horizon of the MBH and
only a final burst of GWs would be emitted, and such burst would be (i) very
difficult to recover, since the very short signal would be buried in a sea of
instrumental and confusion noise and (ii) the information contained in the
signal would be practically nil.  There has been some work on the detectability
of such bursts~\citep{RubboEtAl2006,HopmanFreitagLarson07,YunesEtAl2008,BerryGair2012}, but they would
only be detectable in our galaxy or in the close
neighborhood, but the event rates are rather low, even in the most optimistic
scenarios.

The typical size of the central MBH can be associated with the gravitational
radius (radial horizon location) of the MBH, which in the case of the Milky Way
MBH corresponds to approximately $R^{}_{\rm Schw} \sim 1.3\times 10^{7}$ km
$\approx 4.1 \times 10^{-7}$ pc (neglecting spin contributions).  This number
gives a good indication of how small are these MBHs in size, which means they
have a small cross section and hence, the chances a star has to ``plunge''
through the horizon of the MBH directly are very small.  

To quantify the probability for star absorption by a MBH it is crucial to take
into account the location of the Last Stable Orbit (LSO) of a test massive body
in terms of the MBH parameters. According to General Relativity, these are the mass
$M^{}_{\bullet}$ and its intrinsic angular momentum or spin $S^{}_{\bullet} =
a^{}_{\bullet}\,M^{}_{\bullet} c$, where $a^{}_{\bullet}$ is the spin parameter
with length dimension and subject to the constraints $0\leq
a^{}_{\bullet}c^{2}/(GM^{}_{\bullet}) \leq 1$.  The LSO location is given by a
surface in the orbital configuration space that can be described in terms of
the parameters $(p,e,\iota)$, where $p$ is the dimensionless semilatus rectum,
$e$ is the eccentricity, and $\iota$ is the orbital inclination (with respect
to the MBH spin axis) of the orbit.  Equivalently one can also use the
semimajor axis $a$, or the periapsis location $r^{}_{\rm peri}$, instead of $p$.
The periapsis and apoapsis radii are given then by:

\begin{align}
r^{}_{\rm peri} & = \frac{GM^{}_{\bullet}\,p}{c^{2}(1+e)}\,, \nonumber \\
r^{}_{\rm apo}  & = \frac{GM^{}_{\bullet}\,p}{c^{2}(1-e)}\,. \label{periapo}
\end{align}

It is well-known \citep[see e.g.][]{Bardeen70} that the LSO, for the case
of circular orbits, lies at $3\,R^{}_{\rm Schw}$ for non-spinning MBHs, while
it is shifted out to $9GM^{}_{\bullet}/c^{2}$ for retrograde orbits and down to
$GM^{}_{\bullet}/c^{2}$ for prograde orbits, and these values correspond to the
case of ``extremal'' MBHs, characterized by maximal spins, i.e.
$a^{}_{\bullet}c^{2}/(GM^{}_{\bullet}) = 1\,$.   Despite this fact, traditional
EMRI event rate estimations are based on considerations that neglect the spin
of the MBH.  Taking into account the spin one would expect (considering the
asymmetry between prograde and retrograde orbits) an {\em increase} of the
number of EMRIs since some of the traditionally neglected ``direct plunges''
can actually be disguised EMRIs.

\section{Orbital geodesic motion around a Kerr MBH}
\label{sec.RelMethod}

In order to show the importance of the effect of the spin in the estimation of
the number of EMRIs that will produce a significant amount of GW cycles in the
band of eLISA, we present here two types of calculations.  The first one is to
adapt known results about the stability of orbits of massive objects around a
MBH to our discussion.  The second is an estimation of the number of cycles for
for orbits which would be plunging orbits for a Schwarzschild MBH, or orbits
with no sufficient cycles when the MBH was assumed to be non-spinning for the
case with spin. We show that a significant fraction of them are actually EMRIs
with sufficient cycles to be detected by a space-based observatory like eLISA.
Parts of these calculations, mainly due to the high eccentricities involved,
require numerical computations. 

At this point it is useful to review some basic characteristics of the orbital
geodesic motion of massive bodies around a Kerr MBH.  First of all, the
geometry of a Kerr MBH is axisymmetric (with respect to the spin axis) instead
of spherically-symmetric as in the case of Schwarzschild MBHs, and this means
that the inclination of the orbit with respect to the spin axis, $\iota$, plays
an important role in the dynamics.  Actually, orbits outside the equatorial
plane are not planar, like in Keplerian motion or
orbits around a Schwarzschild MBH, but instead they would precess around a
plane with a certain inclination $\iota$ with a frequency that we call
$f^{}_{\theta}$, where $\theta$ refers to the polar Boyer-Lindquist coordinate
of the MBH~\citep{BoyerLindquist1967,MisnerThorneWheeler1973}.  In addition, relativistic
effects cause precession of the periapsis, and this already happens for
Schwarzschild MBHs, so that we have to consider two more frequencies,
$f^{}_{r}$ and $f^{}_{\varphi}$.  The first one, $f^{}_{r}$, is associated with
the radial motion and the time to go from periapsis, $r^{}_{\rm peri}$, to
apoapsis, $r^{}_{\rm apo}$, and  back.  The second one, $f^{}_{\varphi}$, is
associated with the azimuthal motion around the spin axis and the time to
complete a full turn ($2\,\pi$) around this axis, or in other words, the time
for the azimuthal angle $\varphi$ to increase $2\pi$ radians.  

In summary, generic bound motion around a Kerr MBH exhibits three fundamental
frequencies, $(f^{}_{r},\,f^{}_{\theta},\,f^{}_{\varphi})$ and this implies that
the GW emission of an EMRI will be quite rich in structure (not only these GWs
will contain features with these frequencies but also with a number of harmonics
of them), encoding the detailed geometry of the central Kerr MBH.   The GW
emission will back react on to the system and this translates in particular in
a change of the orbital parameters $(p,\,e,\,\iota)$ of the orbit.  These changes
can be estimated by considering the energy and angular momentum carried away
from the extreme-mass-ratio binary by the GWs emitted.  More specifically, the
GW emission changes the constants of motion of the geodesic motion, namely the
energy per unit mass  (normalized with respect to the star mass, $m$), $E$, the
angular momentum along the spin axis per unit mass, $L^{}_{z}$, and the
so-called Carter constant per unit mass square, $C$, which is associated with
an extra symmetry of the Kerr geometry, similar to what happens in certain
axisymmetric Newtonian potentials~\citep{BT87}.   Actually, the set of
constants $(E,L^{}_{z},C)$ parametrizes the geodesic orbit in the same way as
the set of orbital parameters $(p,e,\iota)$ does.  Therefore, there is a
mapping between these two sets \citep[see][]{Schmidt2002} which is going to
be crucial in the calculations that we present here.  The explicit form of this
mapping is quite complex and we do not include it here, we just mention that we
used the implementation described in~\cite{SopuertaYunes2011}.

In the case of a non-spinning Schwarzschild MBH, where $\iota$ and $C$ do not
play any role, the mapping is much more simple and is given by:

\begin{align}
\frac{E^{2}}{c^{2}} = & \frac{(p-2-2e)(p-2+2e)}{p\,(p-3-e^2)}\,,\\
L^{2}_{z} = & \frac{G^{2}M^2_{\bullet}\,p^2}{c^{2}(p-3-e^2)}\,.
\end{align}

Using the symmetries of the geometry of a Kerr MBH we can separate the equations for
geodesic orbital motion so that the trajectory of a massive body, described in terms
of Boyer-Lindquist coordinates $\{t,r,\theta,\varphi\}$, can be written as follows

\begin{align}
\rho^{2}\, \frac{d{t}}{d\tau} & = \frac{1}{\Delta}\left( \Sigma^{2}\frac{E}{c} - 2 a^{}_{\bullet} 
r^{}_{\bullet} \frac{{L}^{}_{z}}{c} r  \right)  \label{tdot-GR} \\
\rho^{4}\, \left(\frac{d {r}}{d\tau} \right)^{2} & =  \left[ \left( r^{2} + a^{2}_{\bullet} \right) 
\frac{E}{c} - a^{}_{\bullet} \frac{L^{}_{z}}{c} \right]^{2} - \nonumber \\
& \left(\frac{Q}{c^{2}}+r^{2}\right)\Delta  \equiv  R(r) \label{rdot-GR} \\
\rho^{4}\, \left(\frac{d{\theta}}{d \tau} \right)^{2} & = \frac{C}{c^{2}} - \frac{{L}^{2}_{z}}{c^{2}} 
\cot^{2}{\theta} - a^{2}_{\bullet} \left(1 - \frac{{E}^{2}}{c^{2}} \right)\cos^{2}{\theta} \label{thetadot-GR}\\
\rho^{2}\, \frac{d \varphi}{d{\tau}} & = \frac{1}{\Delta}\left[ 2a^{}_{\bullet}r^{}_{\bullet}
\frac{E}{c} r + \frac{{L}^{}_{z}}{c} 
\frac{ \Delta -a^{2}_{\bullet}{\sin^{2}{\theta}}}{\sin^{2}\theta}  \right].
\label{phidot-GR}
\end{align}

\noindent
In the last set of equations we have introduced the following definitions: 

\begin{align}
r^{}_{\bullet} & \equiv \frac{GM^{}_{\bullet}}{c^{2}}, \nonumber \\
Q & \equiv C +  \left(L^{}_{z} - a^{}_{\bullet} E \right)^{2}, \nonumber \\
\rho^{2} & \equiv r^{2} + a^{2}_{\bullet} \cos^{2}{\theta} \nonumber \\
\Delta & \equiv r^{2}-2r^{}_{\bullet}r+a^{2}_{\bullet} = 
r^{2} f + a^{2}_{\bullet}, {\rm ~with~} f \equiv 1 - \frac{2\, r^{}_{\bullet}}{r}. 
\end{align}

\noindent
For convenience, we also define the quantity $\Sigma^{2} \equiv (r^2 +
a^2_{\bullet})^{2} - a^{2}_{\bullet}\Delta\,\sin^{2}\theta\,$.  The first
equation tells us how the coordinate time $t$ changes with respect to the proper
time $\tau$ and the other three describe the trajectory in space.  One can
combine the four equations to obtain the spatial trajectory in terms of
coordinate time $t$, the time of observers at infinity, i.e.
$(r(t),\theta(t),\varphi(t))$.

\section{Kerr and Schwarzschild separatrices}

\begin{figure*}
\resizebox{\hsize}{!}
          {\includegraphics[scale=1,clip]{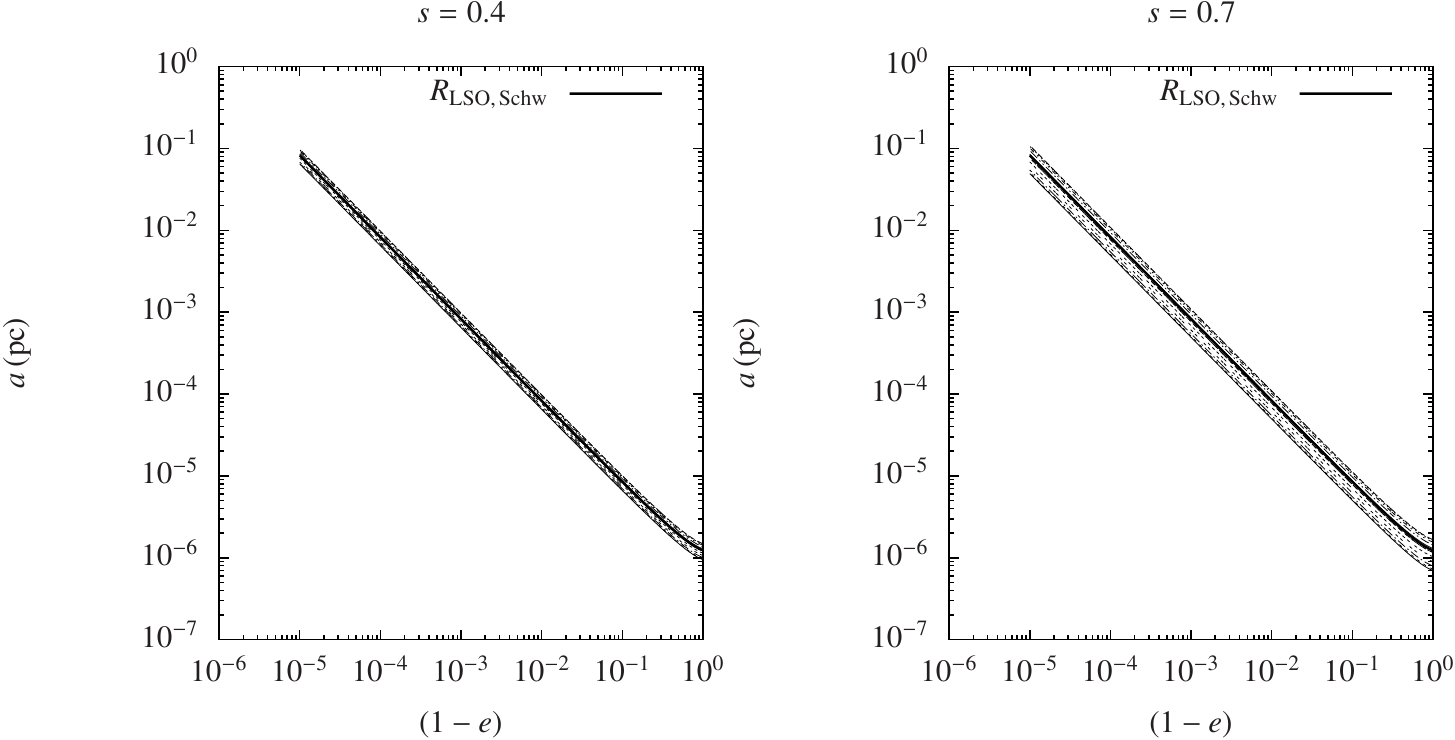}}
\caption
{  
LSO for a MBH of mass $4\times10^4\,M^{}_{\odot}$ and a SBH of mass
$m_{\bullet}=10\,M^{}_{\odot}$ for a Kerr MBH of spin $s=0.4$ (left) and $s=0.7$
(right). The Schwarzschild separatrix is given as a solid black line. Curves
above it correspond to retrograde orbits and inclinations of
$\iota=5.72,\,22.91,\,40.10,\,57.29,\,74.48$ and $89.95^{\circ}$ starting from the
last value ($89.95^{\circ}$). In the left panel we can barely see any difference 
from the different inclinations due to the low value of the spin.
   }
\label{fig.LSO_Spin0p4_0p7}
\end{figure*}

\begin{figure*}
\resizebox{\hsize}{!}
          {\includegraphics[scale=1,clip]{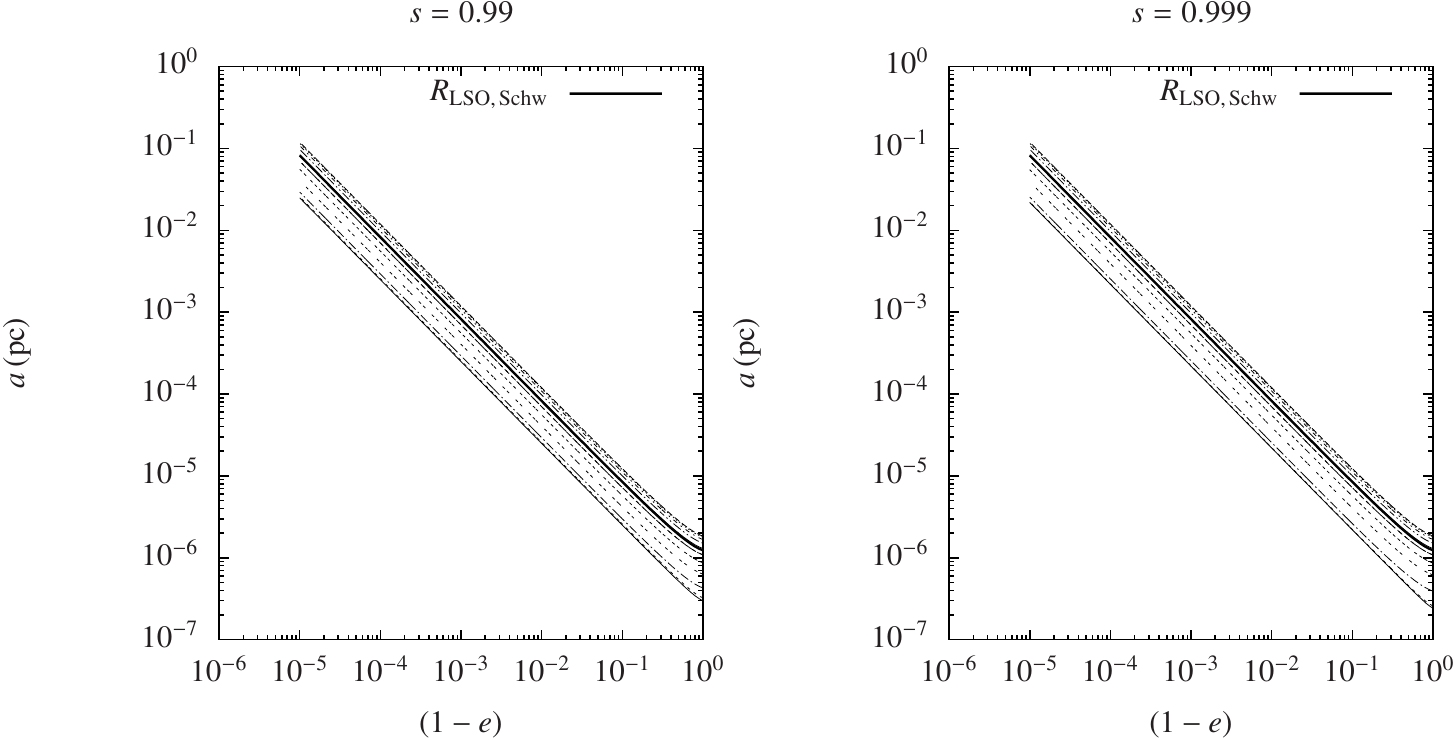}}
\caption
   {  
As in figure \ref{fig.LSO_Spin0p4_0p7} but for a spin of $s=0.99$ (left) and
$s=0.999$ (right panel). The larger the spin, the ``further away'' the Kerr LSO
gets from the Schwarzschild LSO.
   }
\label{fig.LSO_Spin0p99_0p999}
\end{figure*}

In figures~\ref{fig.LSO_Spin0p4_0p7} and~\ref{fig.LSO_Spin0p99_0p999} we show plots
of the location of the LSO in the plane $a$ (pc) - $(1-e)$, including the Schwarzschild
separatrix between stable and unstable orbits, $p -6 - 2e = 0$\footnote{The relation between $p$ and
$a$ is $a = (GM^{}_{\bullet}/c^{2})(p/(1-e^{2}))\,$.}, for both prograde and retrograde orbits
and for different values of the inclination $\iota$.  Each plot corresponds to a different 
value of the spin, showing how increasing the spin makes a difference in shifting the location
of the separatrix between stable and unstable orbits, pushing prograde orbits near 
$GM^{}_{\bullet}/c^{2}$ while retrograde orbits are pushed out towards $9GM^{}_{\bullet}/c^{2}$.  
The procedure we have used to build these plots is a standard one.  Briefly, given a value of 
the dimensionless spin parameter $s\equiv a^{}_{\bullet}c^{2}/(GM^{}_{\bullet})$ and a value 
of the eccentricity $e$ and inclination angle $\iota$ we apply the following algorithm: 

\begin{enumerate}

\item We start from an initial value of the semilatus rectum $p$ which, together with the
value of the eccentricity fixes the apoapsis and periapsis radii through the equations
in~(\ref{periapo}).  On the other hand, the inclination of the orbital plane can also be 
described in terms of the polar angle $\theta$, so that $\theta^{}_{\rm inc}$ represents the
inclination angle and is closely related to $\iota$ \citep[see e.g.,][]{DrascoHughes2004}.
The advantage of using $\theta^{}_{\rm inc}$ is that it is related in a simple way to the 
extrema of $\theta$, $\theta^{}_{\rm min}$ satisfying --see equation~(\ref{thetadot-GR})

\begin{equation}
\left(\frac{d\theta}{d\tau}\right)_{\theta=\theta_{\rm min}} = 0  \,,
\end{equation}

\noindent
by $\theta^{}_{\rm inc} = {\rm sign}(L^{}_{z})\left(\frac{\pi}{2} - \theta^{}_{\rm min}\right)$, with

\begin{equation}
{\rm sign}(L^{}_{z})=
 \begin{cases}
 +1 & \text{for prograde orbits\,,} \\
 -1 & \text{for retrograde orbits\,.}
 \end{cases}
\end{equation}

\noindent
Then, from the condition of extrema of $\theta^{}_{\rm
min}$ and its relation to $\theta^{}_{\rm inc}$, we can find the value of the
Carter constant $C$ in terms of the energy $E$ and angular momentum along the spin
axis $L^{}_{z}$.

\item From the conditions of extrema of periapsis and apoapsis, 

\begin{equation}
\left(\frac{dr}{d\tau}  \right)_{r=r_{\rm peri}}=\left(\frac{dr}{d\tau}\right)_{r=r_{\rm apo}} = 0,
\end{equation}

\noindent
[see equation~(\ref{rdot-GR})], and using the expression of $C$ in terms of $(E,L^{}_{z})$ from
the previous point, we find the values of $(E,L^{}_{z})$ (and hence of $C$ too).

\item The radial motion for geodesic orbits around Kerr has four extrema, the
periapsis and apoapsis locations and two more radii, $r^{}_{3}$ and
$r^{}_{4}$, such that 

\begin{equation}
r^{}_{\rm apo} \geq r^{}_{\rm peri} \geq r^{}_{\rm 3} > r^{}_{4}.  
\end{equation}

\noindent
Actually, $r_{4}$ always lies inside the horizon radius, 

\begin{equation}
r^{}_{\rm H} = \left(\frac{GM^{}_{\bullet}}{c^{2}}\right)\left(1 + \sqrt{1 - s^{2}}\right), 
\end{equation}

\noindent
i.e.  $r^{}_{4}<r^{}_{\rm H}$. For any stable orbit, it is obvious that the
radial motion happens inside the interval $[r^{}_{\rm peri},\,r^{}_{\rm apo}]$.
However, for orbital configurations with $r^{}_{\rm peri} = r^{}_{3}$, the
potential of the MBH changes its shape and the orbits become unstable, marking
the location of the LSO.  In this case, we have that

\begin{equation}
\left(\frac{dR(r)}{dr}\right)_{r=r^{}_{3}} = 0, 
\end{equation}

\noindent
where $R(r)$ denotes the right-hand side of
the evolution equation for the radial position $r$ (see
equation \ref{rdot-GR}). 

\end{enumerate}

\noindent
The calculations in this algorithm are done numerically, so we check whether
this condition is satisfied to some tolerance level.  In the case it is not
satisfied, we use this information to prescribe the next value of $p$ and come back to
the first point in the algorithm.  The process is repeated until we identify
the LSO with the desired accuracy.

\section{Number of cycles}

The second type of relativistic computations that we have performed concerns
the estimation of the number of cycles that certain EMRI orbital configurations
that were thought to be plunging orbits (or orbits with no sufficient cycles) in
the case of non-spinning MBHs can spend in a frequency regime of $f\in
[10^{-4},1]$ Hz during their last year(s) of inspiral before plunging into the
MBH.  This is important to assess how many of these EMRIs will have sufficient 
Signal-to-Noise Ratio (SNR)
to be detectable.   The way in which these estimations have been done is the
following.  We start with a certain orbital configuration characterized by the
orbital parameters $(p^{}_{(0)},e^{}_{(0)},\iota^{}_{(0)})$.  Equivalently, we
can characterise the initial orbital configuration by the constant of motions
$(E^{}_{(0)},L^{}_{z(0)},C^{}_{(0)})$.  Hence, the idea is to track the inspiral
without actually integrating the equations of geodesic motion of
section \ref{sec.RelMethod} or any other type of equations that follow the
trajectory.  Instead, we picture the inspiral as a sequence of geodesic orbits,
each of them characterized by orbital parameters
$(p^{}_{(i)},e^{}_{(i)},\iota^{}_{(i)})$ (or equivalently, constants of motion
$(E^{}_{(i)},L^{}_{z(i)},C^{}_{(i)})$) with $i=0,\ldots,N^{}_{\rm plunge}$,
being $N^{}_{\rm plunge}$ the final plunging configuration.  The transition
between each orbital configuration is governed by the GW emission.  Our
particular algorithm to follow the inspiral goes as follows (our implementation
uses the formul{\ae} in the appendices of~\citealt{SopuertaYunes2011} and the
formulae derived by~\citealt{GG06}): 

\begin{enumerate}

\item Given an orbital configuration $(p^{}_{(i)},e^{}_{(i)},\iota^{}_{(i)})$
and its associated constants of motion $(E^{}_{(i)},L^{}_{z(i)},C^{}_{(i)})$, we
compute the averaged evolution of the constants of motion,
$(\dot{E},\dot{L}^{}_{z},\dot{C})$, using the formul{\ae}
of~\citep{GG06}, which combine post-Newtonian calculations at the 2PN
order with fits to results for the GW emission based on the Teukolsky
formalism~\citep{Teukolsky1973} 
(for details see~\cite{Hughes2000,Hughes2001,GlampedakisEtAl02,DrascoHughes2006,HughesEtAl2005}).

\item For the given orbital parameters we estimate the radial period
$T^{}_{r}$, that is the time to go from apoapsis to periapsis and back to
apoapsis (see~\citealt{Schmidt2002,FujitaHikida2009} for details of this
computation. In~\citealt{FujitaHikida2009} there is a typo in one of the relevant
formulae for our computations fixed in the appendices of~\citealt{SopuertaYunes2011}).

\item We compute the change in the constants of motion.  To that end, due to
the fact that the GW emission by an EMRI is relatively weak, we do not consider
in general just one passage through periapsis, but several of them, say $N^{\rm
peri}_{(i)}$.  Thus, the change in the constants of motion is: 

\begin{align}
(\Delta E^{}_{(i)},\Delta L^{}_{z(i)},\Delta C^{}_{(i)}) & = 
(\dot{E}^{}_{(i)},\dot{L}^{}_{z(i)},\dot{C}^{}_{(i)}) \nonumber \\
&\times N^{\rm peri}_{(i)}\times T^{}_{r}.
\end{align}

And from here,  

\begin{align}
(E^{}_{(i+1)},L^{}_{z(i+1)},C^{}_{(i+1)}) & = (E^{}_{(i)},L^{}_{z(i)},C^{}_{(i)}) +\nonumber \\
& (\Delta E^{}_{(i)},\Delta L^{}_{z(i)},\Delta C^{}_{(i)}).
\label{eq.newconst}
\end{align}

\item From these new constants of motion and using similar techniques to the
ones described above for the LSO computation we can find the new orbital
parameters $(p^{}_{(i+1)},e^{}_{(i+1)},\iota^{}_{(i+1)})$. We then go back to
the first step and follow the inspiral until we detect that the new configuration
corresponds to an unstable orbit corresponding to the plunge.  We estimate the
number of cycles as the number of times that the stellar object turned around
the spin axis, $N^{\,\rm cycles}_{\varphi}$.

\end{enumerate}

In Table~\ref{tab.ncycles} we show some results for a series of inspirals whose initial orbital
parameters are such that the pair $(e,p)$ lies on the separatrix
between stable and unstable orbits in the case of non-spinning MBHs (i.e. $p = 6+2e$).  
Therefore, these are inspirals that under the assumption that spin can be neglected would 
have no cycles in the eLISA band.  However, we can see from the values for the number of cycles
in the table that many of those systems actually perform a significant number of cycles,
more than sufficient to be detectable with good SNR.  The number of cycles has been associated
with $N^{}_{\varphi}$ (the number of times that the azimuthal angle $\varphi$ advances $2\pi$)
which is usual for binary systems.  However, as we have discussed above the structure of 
the waveforms from EMRIs is quite rich since they contain harmonics of three different
frequencies.  Therefore the waveforms have cycles associated with the three frequencies
$(f^{}_{r},f^{}_{\theta},f^{}_{\varphi})$ which makes them quite complex and in principle
this is good for detectability (assuming we have the correct waveform templates).  Another
fact that is worth mentioning is that the cycles quoted in Table~\ref{tab.ncycles} happen
just before plunge and take place in the strong field region very near the MBH horizon. Then,
these cycles should contribute more to the SNR than cycles taking place farther away from the MBH horizon.
Regarding the accuracy of these estimations, the main sources of error are the
approximations made for the radiation-reaction effects, which are based on
a post-Newtonian expansions and fits to results from black hole perturbation
theory.  Corrections from higher-order terms will introduce corrections to
these results that depend on the EMRI configuration, but those corrections
should not affect the magnitude of these numbers. In the Table we quote
the integers that are closer to the numerical result.

\begin{table*}
\begin{center}
\begin{tabular}{|c|c|c|c|c|c|c|c|}
\hline
MBH mass & MBH Spin & Semimajor axis &  Eccentricity  & Inclination & Time to plunge &  Time in band & Cycles in band \\ 
$M^{}_{\bullet}\;(M^{}_{\odot})$ & $s$ & $a^{}_{(0)}$ (pc)  &  $1-e^{}_{(0)}$  &  $\iota^{}_{(0)}$ (rad) & $T^{}_{\rm plunge}$ (yrs) & $T^{}_{\rm band}$ (yrs)  &   $N^{}_{\varphi}$  \\ 
\hline \hline 
$5.0\cdot 10^{4}$ & $0.30$ & $9.57\cdot 10^{-3}$ & $10^{-6}$ & $0.00$ & $3.92\cdot 10^{2}$ & $0.06$ &  $  914$ \\ 
$5.0\cdot 10^{4}$ & $0.30$ & $9.57\cdot 10^{-3}$ & $10^{-6}$ & $0.70$ & $3.92\cdot 10^{2}$ & $0.06$ &  $  625$ \\ 
$5.0\cdot 10^{4}$ & $0.90$ & $9.57\cdot 10^{-3}$ & $10^{-6}$ & $0.00$ & $3.92\cdot 10^{2}$ & $0.11$ &  $ 4174$ \\ 
$5.0\cdot 10^{4}$ & $0.90$ & $9.57\cdot 10^{-3}$ & $10^{-6}$ & $1.00$ & $3.92\cdot 10^{2}$ & $0.08$ &  $ 2646$ \\ 
$1.0\cdot 10^{5}$ & $0.70$ & $1.91\cdot 10^{-2}$ & $10^{-6}$ & $0.00$ & $7.87\cdot 10^{2}$ & $0.29$ &  $ 6968$ \\ 
$1.0\cdot 10^{5}$ & $0.70$ & $1.91\cdot 10^{-2}$ & $10^{-6}$ & $1.00$ & $7.86\cdot 10^{2}$ & $0.23$ &  $ 3411$ \\ 
$1.0\cdot 10^{5}$ & $0.99$ & $1.91\cdot 10^{-2}$ & $10^{-6}$ & $0.00$ & $4.70\cdot 10^{3}$ & $0.38$ &  $ 8938$ \\ 
$1.0\cdot 10^{5}$ & $0.99$ & $1.91\cdot 10^{-2}$ & $10^{-6}$ & $0.70$ & $3.92\cdot 10^{3}$ & $0.32$ &  $ 7892$ \\ 
$5.0\cdot 10^{5}$ & $0.30$ & $9.57\cdot 10^{-2}$ & $10^{-6}$ & $0.00$ & $4.31\cdot 10^{4}$ & $1.98$ &  $ 8246$ \\ 
$5.0\cdot 10^{5}$ & $0.30$ & $9.57\cdot 10^{-2}$ & $10^{-6}$ & $1.00$ & $4.31\cdot 10^{4}$ & $1.61$ &  $ 3061$ \\ 
$5.0\cdot 10^{5}$ & $0.95$ & $9.57\cdot 10^{-2}$ & $10^{-6}$ & $0.00$ & $5.10\cdot 10^{4}$ & $2.00$ &  $40093$ \\ 
$5.0\cdot 10^{5}$ & $0.95$ & $9.57\cdot 10^{-2}$ & $10^{-6}$ & $1.00$ & $4.31\cdot 10^{4}$ & $2.00$ &  $27600$ \\ 
$1.0\cdot 10^{6}$ & $0.30$ & $1.91\cdot 10^{-2}$ & $10^{-5}$ & $0.00$ & $1.27\cdot 10^{4}$ & $1.98$ &  $10943$ \\ 
$1.0\cdot 10^{6}$ & $0.30$ & $1.91\cdot 10^{-2}$ & $10^{-5}$ & $1.00$ & $1.19\cdot 10^{4}$ & $1.91$ &  $ 3552$ \\ 
$1.0\cdot 10^{6}$ & $0.70$ & $1.91\cdot 10^{-2}$ & $10^{-5}$ & $0.00$ & $1.35\cdot 10^{4}$ & $1.99$ &  $51308$ \\ 
$1.0\cdot 10^{6}$ & $0.70$ & $1.91\cdot 10^{-2}$ & $10^{-5}$ & $1.00$ & $1.20\cdot 10^{4}$ & $1.99$ &  $23291$ \\ 
$1.0\cdot 10^{6}$ & $0.90$ & $1.91\cdot 10^{-2}$ & $10^{-5}$ & $0.00$ & $1.40\cdot 10^{4}$ & $1.99$ &  $58841$ \\ 
$1.0\cdot 10^{6}$ & $0.90$ & $1.91\cdot 10^{-2}$ & $10^{-5}$ & $1.00$ & $1.17\cdot 10^{4}$ & $2.00$ &  $38245$ \\ 
$1.0\cdot 10^{6}$ & $0.99$ & $1.91\cdot 10^{-2}$ & $10^{-5}$ & $0.00$ & $1.43\cdot 10^{4}$ & $2.00$ &  $61726$ \\ 
$1.0\cdot 10^{6}$ & $0.99$ & $1.91\cdot 10^{-2}$ & $10^{-5}$ & $1.00$ & $1.17\cdot 10^{4}$ & $2.00$ &  $47678$ \\ 
$5.0\cdot 10^{6}$ & $0.30$ & $9.57\cdot 10^{-2}$ & $10^{-5}$ & $0.00$ & $1.44\cdot 10^{5}$ & $1.93$ &  $ 5258$ \\ 
$5.0\cdot 10^{6}$ & $0.30$ & $9.57\cdot 10^{-2}$ & $10^{-5}$ & $1.00$ & $1.36\cdot 10^{5}$ & $0.00$ &  $    0$ \\ 
$5.0\cdot 10^{6}$ & $0.70$ & $9.57\cdot 10^{-2}$ & $10^{-5}$ & $0.00$ & $1.55\cdot 10^{5}$ & $2.00$ &  $40687$ \\ 
$5.0\cdot 10^{6}$ & $0.70$ & $9.57\cdot 10^{-2}$ & $10^{-5}$ & $1.00$ & $1.36\cdot 10^{5}$ & $2.00$ &  $14936$ \\ 
$5.0\cdot 10^{6}$ & $0.90$ & $9.57\cdot 10^{-2}$ & $10^{-5}$ & $0.00$ & $1.61\cdot 10^{5}$ & $2.00$ &  $41369$ \\ 
$5.0\cdot 10^{6}$ & $0.90$ & $9.57\cdot 10^{-2}$ & $10^{-5}$ & $1.00$ & $1.35\cdot 10^{5}$ & $1.99$ &  $30695$ \\ 
$1.0\cdot 10^{7}$ & $0.30$ & $1.91$ & $10^{-6}$ & $0.00$ & $4.02\cdot 10^{6}$ & $1.99$ &  $ 3089$ \\ 
$1.0\cdot 10^{7}$ & $0.30$ & $1.91$ & $10^{-6}$ & $1.00$ & $3.78\cdot 10^{6}$ & $0.00$ &  $    0$ \\ 
$1.0\cdot 10^{7}$ & $0.70$ & $1.91$ & $10^{-6}$ & $0.00$ & $4.27\cdot 10^{6}$ & $2.00$ &  $23425$ \\ 
$1.0\cdot 10^{7}$ & $0.70$ & $1.91$ & $10^{-6}$ & $1.00$ & $3.79\cdot 10^{6}$ & $1.98$ &  $ 8747$ \\ 
$1.0\cdot 10^{7}$ & $0.99$ & $1.91\cdot 10^{-1}$ & $10^{-5}$ & $0.00$ & $1.44\cdot 10^{6}$ & $1.98$ &  $22455$ \\ 
$1.0\cdot 10^{7}$ & $0.99$ & $1.91\cdot 10^{-1}$ & $10^{-5}$ & $1.00$ & $1.18\cdot 10^{6}$ & $1.99$ &  $28589$ \\ 
$5.0\cdot 10^{7}$ & $0.30$ & $9.57\cdot 10^{-1}$ & $10^{-5}$ & $0.00$ & $1.44\cdot 10^{7}$ & $0.00$ &  $    0$ \\ 
$5.0\cdot 10^{7}$ & $0.30$ & $9.57\cdot 10^{-1}$ & $10^{-5}$ & $1.00$ & $1.36\cdot 10^{7}$ & $0.00$ &  $    0$ \\ 
$5.0\cdot 10^{7}$ & $0.70$ & $9.57\cdot 10^{-1}$ & $10^{-5}$ & $0.00$ & $1.55\cdot 10^{7}$ & $1.72$ &  $ 4247$ \\ 
$5.0\cdot 10^{7}$ & $0.70$ & $9.57\cdot 10^{-1}$ & $10^{-5}$ & $1.00$ & $1.35\cdot 10^{7}$ & $0.00$ &  $    0$ \\ 
$5.0\cdot 10^{7}$ & $0.99$ & $9.57\cdot 10^{-1}$ & $10^{-5}$ & $0.00$ & $1.65\cdot 10^{7}$ & $1.88$ &  $ 4422$ \\ 
$5.0\cdot 10^{7}$ & $0.99$ & $9.57\cdot 10^{-1}$ & $10^{-5}$ & $1.00$ & $1.35\cdot 10^{7}$ & $1.52$ &  $ 4625$ \\ 
\hline \hline 
\end{tabular}
\end{center}
\caption{This table shows the main properties of some (prograde) inspirals that initially lie on the separatrix (LSO) of non-spinning MBHs
and hence they would not be detectable in the eLISA band.  The numbers in the first five columns have been already introduced in
the text. The sixth column gives the time it takes for each inspiral to get to plunge.  The seventh column shows how much time
it spends in band assuming the plunge occurs at the end of the eLISA mission time (assumed to be $2\,$ yrs here).  The last
column show the number of orbital cycles in band (during $T^{}_{\rm band}$), defined as the number of times that the azimuthal angle $\varphi$ advances $2\pi$ 
during the last two years before plunge. The number of GW cycles can be then defined as twice this number.}
\label{tab.ncycles}
\end{table*}

\section{Impact on event rates}
\label{sec.eventrates}

Only a certain fraction of
stars in phase space will come close enough to interact with the MBH.
These stars are said to belong to the ``loss-cone'' \citep[see e.g.][]{FR76,AS01}.

For radii larger than $0.01$ pc the main leading mechanism for producing EMRIs
is two-body relaxation
\citep{HopmanAlexander05,Amaro-SeoaneEtAl07,Amaro-SeoaneLRR2012}, and this is
the region of phase-space in which our analysis is applied with priority, since
for a Schwarzschild MBH one has just direct plunges. For radii below $0.01$ pc
we note that the enhancement in the EMRI event rate due to resonant relaxation predicted by
\cite{HopmanAlexander05} is severely affected by the presence of a blockade in
the rate at which orbital angular momenta change takes place. This so-called
``Schwarzschild barrier'' is a result of the impact of relativistic precession
on to the stellar potential torques, as recently shown by \cite{MerrittEtAl11}
with a few direct-summation $N-$body simulations expanded with a statistical
Monte-Carlo study. Indeed, this ``Schwarzschild barrier'' has been corroborated
in an independent work with a statistical study based on a sample of {\em some
2,500} direct-summation $N-$body simulations by
\cite{BremAmaroSeoaneSopuerta2012} in which the authors include post-Newtonian
corrections and also, for the first time, the implementation of a solver of
geodesic equations in the same code.  This barrier poses a real problem for the
production of low-eccentricity EMRIs. However, high-eccentricity EMRIs, which
had been classified until now wrongly of ``plunges'' do not have this problem,
since they are a product of pure two-body relaxation. This is why we will only
focus on two-body relaxation for the estimation of the rates.

The event rate can be hence approximately calculated as

\begin{equation}
\dot{N}_{\rm EMRI} \simeq \int^{a_{\rm EMRI}}_{0} \frac{dN_{\bullet}(a)}
                     {\ln{\left(\theta_{\rm LC}^{-2}\right)}\,t^{}_r(a)}\,,
\label{eq.Ndot}
\end{equation}

\noindent
with $\theta_{\rm LC}$ the loss-cone angle, 
$N^{}_{\bullet}(a)$ the number of stellar black holes (SBHs) within a 
semi-major axis $a$ and $a^{}_{\rm EMRI}$
a maximum radius within which we estimate the event rate $\dot{N}^{}_{\rm EMRI}$.
We assume that the SBHs distribute around the central MBH following a power-law
cusp of exponent $\gamma$, i.e. that the density profile follows $\rho \propto r^{-\gamma}$
within the region where the gravity of the MBH dominates the gravity of the stars, with
$\gamma$ ranging between 1.75 and 2 for the heavy stellar components
\citep{Peebles72,BW76,BW77,ASEtAl04,PretoMerrittSpurzem04,AlexanderHopman09,PretoAmaroSeoane10,Amaro-SeoanePreto11}
and see \cite{Gurevich64} for an interesting first idea of this concept\footnote{The
authors obtained a similar solution for how electrons distribute around a
positively charged Coulomb centre.}. We then have that the number of stars within a radius
$r$ is

\begin{equation}
n(r) = \frac{\left(3-\gamma\right)}{4\,\pi} \left(\frac{M^{}_{\bullet}}{m^{}_{\star}}\right) 
       \left(\frac{1}{R_{\rm infl}^3} \right) \left( \frac{r}{R^{}_{\rm infl}}  \right)^{-\gamma}\,. 
\label{eq.ninr}
\end{equation}

\noindent
Hence, the number of SBHs within $a$ is

\begin{equation}
N^{}_{\bullet}(a) \simeq N^{}_{0}\left(\frac{a}{R^{}_{0}}\right)^{3-\gamma}\,.
\label{eq.Nbullet}
\end{equation}

\noindent
Hence, we have that

\begin{equation}
dN^{}_{\bullet}(a) = (3-\gamma)\frac{N^{}_{0}}{R^{}_{0}}
                  \left(\frac{a}{R_{0}}\right)^{2-\gamma}da \,.
\end{equation}

\subsection{The Schwarzschild case}

\noindent
We know that \citep[see e.g.][]{AL01,AS01} $\theta^{-2}_{\rm LC} \simeq J^{}_{\rm max}/J^{}_{\rm LC}$. 
Since the loss-cone angular momentum can be approximated as 
$J^{}_{\rm LC} \simeq 4\,G M^{}_{\bullet}/c$ and $J^{}_{\rm max}=\sqrt{G M^{}_{\bullet}a}$, we have that

\begin{equation}
\theta^{-2}_{\rm LC} \simeq \sqrt{\frac{a}{8\,R_{\rm Schw}}},
\label{eq.thetaLC2}
\end{equation}

\noindent
We assume also that relaxation is dominated by a single stellar black hole (SBH)
population, since because of mass segregation the most massive objects sink down to the centre and
the light stars are pushed out from the centre.
The relaxation time at a distance $a$ is

\begin{equation}
t^{}_{r}(a) = t^{}_{0}\left(\frac{a}{R^{}_{0}}\right)^{\gamma-3/2}\,,
\label{eq.Tr}
\end{equation}

\noindent
with

\begin{equation}
t^{}_{0} = 0.3389\, \frac{\sigma_{0}^3}{\ln \Lambda\,G^2m_{\bullet}^2\,n^{}_{0}}\,.
\label{eq.t0}
\end{equation}

\noindent
Since \citep[see e.g.][]{Amaro-SeoaneEtAl07}

\begin{align}
n^{}_{0}     & = \frac{3-\gamma}{4\pi}{N^{}_{0}}{R_{0}^3} \\
\sigma_{0}^2 & = \frac{1}{1+\gamma}\frac{G M^{}_{\bullet}}{R^{}_{0}},
\end{align}

\noindent
we have that equation~(\ref{eq.t0}) becomes

\begin{equation}
t^{}_{0} \simeq 4.26 \, \frac{1}{(3-\gamma)(1+\gamma)^{3/2}}
             \frac{\sqrt{R_{0}^3(G M^{}_{\bullet})^{-1}}}{\ln \Lambda\,N^{}_{0}}
             \left(\frac{M^{}_{\bullet}}{m^{}_{\bullet}}\right)^2 \,.
\label{eq.t0final}
\end{equation}

\noindent
We now define a {\em critical} radius at
which the two regimes that lead the evolution of the EMRI decouple. The first
evolution is dominated by relaxational processes, via exchange of $E$ and $J$
between SBHs on a capture orbit with stars from the surrounding stellar system,
while in the second regime the evolution of the EMRI is totally dominated by
the emission of gravitational waves. This is given in figure 1 of
\cite{Amaro-SeoaneEtAl07} with their red curve. In other words, the line gives us
the radius as a function of $a$ at which the relaxational time at periapsis is approximately
equal to the timescale defined by the approximation of \cite{Peters64}. Hence,
we have to solve

\begin{align}
t_{\rm r}(a)(1-e) & = K\,t_{\rm GW}(a,\,e) \nonumber \\
J(a,\,e)          & = J_{\rm LC} \nonumber \\
(1-e)\,a          & = \frac{8\,G{\cal M}_{\bullet}}{c^2}
\label{eq.Trelperi}
\end{align}

\noindent
In the first equality, $K$ is a factor of order unity.  
In the last equality we assume a Schwarzschild radius and we assume that the LSO
is at $4 \times\,R_{\rm Schw}$
Approximating $e \sim
1$, the function $f(e)$ from \cite{Peters64} $f(e) = 425/(768\,\sqrt{2})$.
Hence

\begin{equation}
t^{}_{\rm GW}(a,\,e) \cong \sqrt{2}\,\frac{24}{85}\frac{c^5}{G^3}
                        \frac{a^4}{M_{\bullet}^2\,m^{}_{\bullet}}
                        \left(1-e\right)^{7/2}\,.
\label{eq.Tgwae}
\end{equation}

\noindent
And so, finally from equations~(\ref{eq.Trelperi}), (\ref{eq.Tr}) and~(\ref{eq.t0final}) 
and solving for $a^{}_{\rm EMRI}$, we have that

\begin{equation}
a^{}_{\rm EMRI} \simeq R^{}_{0} \left[ 16.97\,K\,\left(3-\gamma\right)  \left(1+\gamma\right)^{3/2}
\ln \Lambda\,N^{}_{\bullet} \frac{m^{}_{\bullet}}{M^{}_{\bullet}}\right]^{\frac{1}{\gamma -3}} \,. 
\label{eq.aEMRI}
\end{equation}

\noindent
Or, absorbing some constants into a newly defined $K^{}_{\gamma}$,

\begin{align}
a^{}_{\rm EMRI} & = K^{}_{\gamma}\,R^{}_{0}\,\left(\frac{1}{\ln \Lambda}\,
                 \frac{M^{}_{\bullet}}{N^{}_{\bullet}\,m^{}_{\bullet}}\right)^{\frac{1}{3-\gamma}} \nonumber \\
K^{}_{\gamma}   & := \left[ 16.97\,K\,\left(3-\gamma\right)\left(1+\gamma\right)^{3/2}\right]\,.
\end{align}

\noindent
We can then derive the event rate for the Schwarzschild case, based on \ref{eq.Ndot},

\begin{align}
\dot{N}_{\rm EMRI}^{\rm Schw} & \cong 0.235\,\frac{4\,\left(3-\gamma\right)^2
\left(1+\gamma\right)^{3/2}}{9-4\gamma}\,K_{\gamma}^{\frac{9-4\gamma}{2}}\nonumber \\
& \frac{\ln \Lambda^{\frac{2\gamma-3}{6-2\gamma}}}{\ln \left(\frac{a^{}_{\rm EMRI}}{8\,R^{}_{\rm Schw}}\right)}\,
\left( \frac{N^{}_{\bullet}m^{}_{\bullet}}{M^{}_{\bullet}} \right)^{\frac{3}{6-2\gamma}}
\sqrt{\frac{G M^{}_{\bullet}}{R_{0}^3}}\,.
\label{eq.NemriFinal}
\end{align}

\noindent
The last equation is based on the assumption that equation~(\ref{eq.Trelperi}) holds, and
the last equality, $(1-e)a=8\,G M^{}_{\bullet}/c^2$, is the ``effective'' value of
the periapsis for the last parabolic stable orbit (LSO from now onwards) for a
Schwarzschild MBH. I.e. if the star is on an orbit that in Newtonian dynamics
leads to a periapsis smaller than this value, the star disappears if we take
into account relativistic dynamics.

\subsection{The Kerr case}

In the Kerr case we simply have to recalculate where this LSO is by taking into
account the value of the spin of the MBH. We then have to either shrink or
enlarge it by a certain factor function of the inclination and spin, 
${\cal W}(\iota,\,s)$, so that the effective pericentre of LSO is

\begin{equation}
(1-e)\,a = {\cal W}(\iota,\,s) \times \frac{8\,G M^{}_{\bullet}}{c^2}\,.
\label{eq.DefW}
\end{equation}

\noindent
This quantity can be derived from the separatrices of the figures in section
\ref{sec.RelMethod}.  If the orbit can get closer to the MBH in the Kerr case,
then ${\cal W}(\iota,\,s)<1$; otherwise ${\cal W}(\iota,\,s)>1$. 
Since the separatrices are nearly parallel, 
we hence can define ${\cal W}(\iota,\,s)$ like

\begin{equation}
{\cal W}(\iota,\,s) := \left\langle\frac{a^{}_{\rm LSO,\,Kerr}}{a^{}_{\rm LSO,\,Schw}} \right\rangle = 
         \frac{1}{N}\,\sum_{i} \frac{a^{}_{\rm LSO,\,Kerr}(e^{}_{i})}{a^{}_{\rm LSO,\,Schw}(e^{}_{i})}\,.
\label{eq.w}
\end{equation}

\noindent
That is, given a spin and an inclination, we take for $N$ values of the
eccentricity the semimajor axis corresponding to the Kerr value, $a^{}_{\rm
LSO,\,Kerr}(e)$ and to the Schwarzschild case, $a^{}_{\rm LSO,\,Schw}$ and we sum
for the ratio of the two semi-majors over all eccentricities. This allows us to
calculate by how much the LSO has been ``shifted''.

\begin{table}
\begin{center}
\begin{tabular}{| c | c | c |}
\hline
Spin ($s$) & Inclination ($\iota$, rad)  &  ${\cal W}(\iota,\,s)$         \\
\hline \hline 
         0.900          &         0.100       &      0.429452\\
         0.900          &         0.400       &      0.448093\\
         0.900          &         0.700       &      0.499450\\
         0.900          &         1.000       &      0.598278\\
         0.900          &         1.300       &      0.739339\\
         0.900          &         1.570       &      0.883679\\
         \hline
         0.900          &        -0.100       &      1.415955\\
         0.900          &        -0.400       &      1.377239\\
         0.900          &        -0.700       &      1.295011\\
         0.900          &        -1.000       &      1.175760\\
         \hline\hline
         0.950          &         0.100       &      0.370036\\
         0.950          &         0.400       &      0.386009\\
         0.950          &         0.700       &      0.436921\\
         0.950          &         1.000       &      0.548352\\
         0.950          &         1.300       &      0.708257\\
         0.950          &         1.570       &      0.867320\\
         \hline
         0.950          &        -0.400       &      1.396449\\
         0.950          &        -0.700       &      1.309052\\
         0.950          &        -1.000       &      1.181942\\
         0.950          &        -1.300       &      1.024866\\
         \hline\hline
         0.990          &         0.100       &      0.297301\\
         0.990          &         0.400       &      0.306924\\
         0.990          &         0.700       &      0.354716\\
         0.990          &         1.000       &      0.494738\\
         0.990          &         1.300       &      0.679468\\
         0.990          &         1.570       &      0.852821\\
         \hline
         0.990          &        -0.100       &      1.454732\\
         0.990          &        -0.400       &      1.411720\\
         0.990          &        -0.700       &      1.320145\\
         0.990          &        -1.000       &      1.186631\\
         0.990          &        -1.300       &      1.020814\\
         \hline\hline
         0.999          &         0.100       &      0.260205\\
         0.999          &         0.400       &      0.264063\\
         0.999          &         0.700       &      0.310302\\
         0.999          &         1.000       &      0.479038\\
         0.999          &         1.300       &      0.672349\\
         0.999          &         1.570       &      0.849364\\
         \hline
         0.999          &        -0.100       &      1.458589\\
         0.999          &        -0.400       &      1.415145\\
         0.999          &        -0.700       &      1.322624\\
         0.999          &        -1.000       &      1.187655\\
         0.999          &        -1.300       &      1.019828\\
\hline \hline
\end{tabular}
\end{center}
\caption{Values for ${\cal W}$.}
\label{tab.W}
\end{table}

If we redo the derivation of~(\ref{eq.NemriFinal}) taking into account equation
\ref{eq.DefW}, we can obtain $\dot{N}^{}_{\rm EMRI}$ in function of ${\cal W}$,
$\dot{N}_{\rm EMRI}^{\rm Kerr}$ and, hence, we can calculate the ratio of the
rates as a function of the inclination $\iota$ and spin $s$:

\begin{align}
{{a}_{\rm EMRI}^{\rm Kerr}} & =  {{a}_{\rm EMRI}^{\rm Schw}} \times 
{\cal W}^{\frac{-5}{6-2\gamma}}(\iota,\,s)\\
{\dot{N}_{\rm EMRI}^{\rm Kerr}} & =  {\dot{N}_{\rm EMRI}^{\rm Schw}} \times 
{\cal W}^{\frac{20\gamma-45}{12-4\gamma}} (\iota,\,s) \,.
\label{eq.NAEMRIW}
\end{align}

\noindent
For instance, for a spin of $s=0.999$ and an inclination of $\iota = 0.4\,$rad, we
estimate that ${\cal W}\sim 0.26$ and, thus, $\dot{N}_{\rm EMRI}^{\rm Kerr}
\sim 30$. I.e. {\em we boost the event rates by a factor of 30} in comparison to a
non-rotating MBH.

\section{Net effect of resonant relaxation on EMRI rates}

\subsection{The role of vectorial resonant relaxation}

To understand the impact of the previous calculation on event rates for EMRIs
we have to elaborate on prograde and retrograde orbits.  We have seen that
retrograde EMRI orbits see the central MBH as an effectively ``larger'' MBH;
i.e. it is easier to plunge through the horizon. Contrary, prograde EMRI orbits
``see'' the effective size of the MBH shrink and, thus, have a harder time in
hitting the central MBH.  It is therefore important to assess the orientation
of orbits in the regime of interest. 
It takes on average  (vectorial) resonant relaxation (RR)
a time $t^{}_{\rm RR,\,v}$ to rotate coherently the orbital plane of
an orbit by an angle $\pi/2$ \citep{HopmanAlexander06}.
To change a prograde (retrograde) orbit to a retrograde (prograde) orbit, it
takes four times longer: The $\pi/2$ rotation is the maximum that can be
obtained over the self-quenching time; the rest to get up to a full $\pi$
rotation is done non-coherently over 4 coherence times \citep[see][for a
discussion of the numeric prefactors]{BregmanAlexander11}.

It should be noted that vector RR is invariant under precession
\citep[see e.g.][]{HopmanAlexander06}. We must note also that the change in the
inclination of the orbit with respect to the spin axis due to GW emission is
relatively rather small (see~\cite{Hughes2000,Hughes2001}), so small that
frequently it has been assumed to be constant, which provides an extra equation
for the evolution of the Carter constant in the inspiral process, making things
significantly simpler.

The dependence of the transverse RR torque (i.e. direction-changing torque) on
the eccentricity has been measured from simulations by \cite{GurkanHopman07}.
In their work, the authors derive that it grows quadratically by a factor $3$ in
total from $0$ to $1$.

The radius of the sphere of influence is

\begin{equation}
r^{}_{\rm infl} = \frac{1}{1+\gamma}\frac{G M^{}_{\bullet}}{\sigma_0^2}\approx 
1~{\rm pc}\frac{1}{1+\gamma}\left(\frac{M^{}_{\bullet}}{10^6\,M^{}_\odot}\right)
\left(\frac{60~{\rm km/s}}{\sigma_0}\right)^2\,,
\end{equation}

\noindent
for a given exponent $\gamma$.   Within this radius the relaxation time is

\begin{align}
t^{}_{\rm r}(r) & \propto \left(1+\gamma \right)^{-3/2}\frac{\ln\Lambda\,\sigma^3(r)}{G^2\langle m\rangle
m^{}_{\rm CO}n(r)} \nonumber \\
& \simeq 2\times 10^8\,{\rm yr}\left(1+\gamma \right)^{-3/2}\left(\frac{\sigma}{100\,\kms}\right)^3
\left(\frac{10\,\Msun}{m^{}_{\rm CO}}\right) \nonumber \\
& \left(\frac{10^6\,\Msun{\rm pc}^{-3}}{\langle m\rangle n}\right)\,.
\label{eq.Trlx}
\end{align}

\noindent
In this equation we follow the usual notation: $\sigma(r)$ is the local velocity
dispersion; for $r<r^{}_{\rm infl}$ it is approximately the Keplerian orbital
speed $\sqrt{G M^{}_{\bullet} r^{-1}}$; $n(r)$ is the {\em local} density of stars,
$\langle m\rangle$ is the average stellar mass, $m^{}_{\rm CO}$ is the individual mass of the
compact object, which we assume to be all SBHs, and take a mass of $m^{}_{\rm
CO}=10\,M^{}_\odot$ for all of them. In the vicinity of a MBH, ($r<r^{}_{\rm infl}$),
$\Lambda \approx M^{}_{\bullet} / m^{}_{\star}$ \citep{BW76,LS77}, and typically $\ln\Lambda\sim 11$.

Relaxation redistributes orbital energy amongst stellar-mass objects until SBHs
form a power-law density cusp, $n(r)\propto r^{-\gamma}$ with $\gamma \simeq
1.75$ around the MBH, while less massive species arrange themselves into a
shallower profile, with $\alpha \simeq 1.4-1.5$ as we have mentioned earlier,
although recent studies have found a general solution to the problem of mass
segregation around MBH in galactic nuclei, with a more efficient diffusion for
the heavy stars, reaching a $\gamma \sim -2$ in the ``strong mass-segregation''
regime \citep{AlexanderHopman09,PretoAmaroSeoane10,Amaro-SeoanePreto11}.

\noindent
Since $\sigma(r)^2 = G M^{}_{\bullet} /r$ and we take that $\langle
m\rangle = 0.7\,M^{}_{\odot}$ and, as mentioned, $m^{}_{\rm CO}=10\,M^{}_{\odot}$, we
have all information to derive $t^{}_{\rm r,\,peri}(r):=(1-e)\,t^{}_{\rm
r}$ from Equation~(\ref{eq.Trlx}) and (\ref{eq.ninr}).

As regards the explicit expression for the 
characteristic timescale for {\em vectorial} resonant relaxation, from
\cite{HopmanAlexander06} we have that 

\begin{equation}
t^{}_{\rm RR,\,v} = P(a) \frac{M^{}_{\bullet}}{m^{}_{\star}} \frac{\beta_{v}(e)^2}{\sqrt{N(a)}}
\,,\label{eq.}
\end{equation}
where we have taken into account the corrections for high values of $e$ as
given in \cite{GurkanHopman07}, $\beta^{}_{v}(e)=0.28\,(e^2+0.5)$ and $P(a) =
a^{3/2}/(G M^{}_{\bullet})^{1/2}$. This allows us to follow the dependence with
the radius (and eccentricity) of the ratio $t^{}_{\rm r,\,peri}/t^{}_{\rm RR,\,v}$.

\begin{figure*}
\resizebox{\hsize}{!}
          {\includegraphics[scale=1,clip]{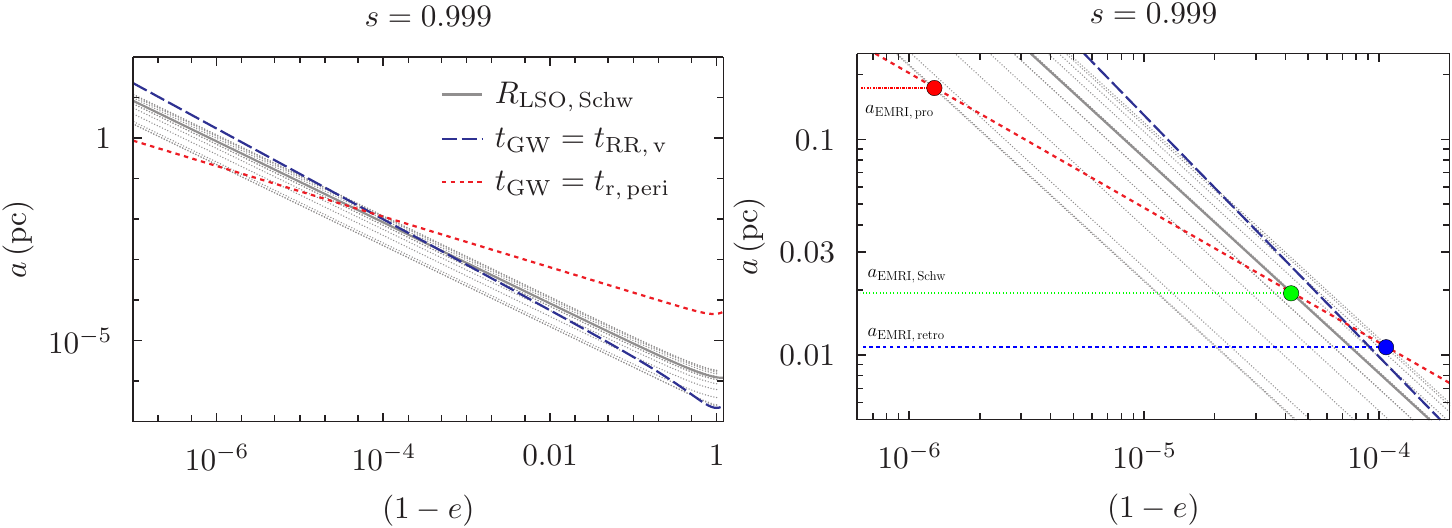}}
\caption{
{\em Left panel:}
Relation between different timescales in the $s=0.999$ case. As in the prior section,
we display the Schwarzschild separatrix as a solid, black line and the separatrices
for different inclinations with different curves in light grey. The dashed, blue line
shows the value of $a$ and $1-e$ for which the vectorial resonant relaxation timescale
($t^{}_{\rm RR,\,v}$) is equal to the gravitational loss timescale ($t^{}_{\rm GW}$). The
dashed, dotted line corresponds to the values of $a$ and $1-e$ for which the relaxation
time at periapsis ($t^{}_{\rm r,\,peri}$) equals the gravitational loss timescale.
{\em Right panel: }
Same as the left panel but zoomed to see where the dashed, red curve intersects the last
one of the retrograde, Schwarzschild and prograde separatrices. We show this with a blue
dot and a long-dashed curve for the retrograde case which yields the last separatrix,
with a green dot and a green, short-dashed curve for the Schwarzschild separatrix and with
a red dot and a dash-dotted curve for the last separatrix of the prograde case. These
lines give us $a_{\rm EMRI,\,retro}$, $a_{\rm EMRI,\,Schw}$ and
$a_{\rm EMRI,\,pro}$, correspondingly.
   }
\label{fig.Tgw_TRRv_Trlx}
\end{figure*}

If we now equate the timescales of interest, the gravitational radiation driven
time $t^{}_{\rm GW}$, defined as in the approximation of \cite{Peters64}, to the
two-body relaxation time at periapsis, $t^{}_{\rm r,\,peri}$, we obtain the
short-dashed curve of figure~\ref{fig.Tgw_TRRv_Trlx} on the left of this line,
the contribution of GW radiation to orbital evolution dominates over two-body
relaxation. In the absence of resonant relaxation, if a SBH crosses this line
from the right (lower eccentricities), it will become an EMRI, provided, of
course, that it is still on a stable orbit, i.e.\ above the separatrix
corresponding to its orbital orientation.  For a Schwarzschild MBH, all
separatrices are the same and there is a unique critical point (PS). A SBH with
a semi-major axis larger that the value at PS will experience a direct plunge
if relaxation brings its eccentricity to a high value because it will cross the
separatrix (and be swallowed in less than an orbital time) before it has a
chance to enter into the GW-dominated regime.  Conversely, objects with smaller
semi-major axis values are much more likely to end up as EMRIs rather than
plunges.

For a fast spinning SMBH, the separatrix for prograde orbits is shifted to
significantly lower $a$ values, with a corresponding higher value of the
critical semi-major axis, corresponding to the point PP in
figure~\ref{fig.Tgw_TRRv_Trlx}.  As we have explained above, it is this effect
which can lead to a significant increase in the EMRI rate, combined with the
fact that the critical point for retrograde orbits (PR) is much less affected
and that an isotropic orbit distribution is expected, thanks to relaxational
processes.  However this increase in EMRI rate would can be thwarted by vector
RR if this process can change the orbital orientation of a SBH after it has
crossed the ``$t^{}_{\rm GW}=t^{}_{\rm r,\,peri}$'' line and before it has completed
its GW-driven inspiral, i.e.\ on a timescale shorter than $t^{}_{\rm GW}$. Indeed,
if the orbit becomes significantly less prograde as the the inspiral takes
place, due to RR, the separatrix moves up and the SBH might suddenly find
itself on a plunge orbit.

To check for this possibility, we also plot, in figure~\ref{fig.Tgw_TRRv_Trlx},
a long-dashed line corresponding to the condition $t^{}_{\rm GW}=t^{}_{\rm RR,\,v}$,
with $t^{}_{\rm GW}<t^{}_{\rm RR,\,v}$ on the left of this line. SBHs that cross the
``$t^{}_{\rm GW}=t^{}_{\rm r,\,peri}$'' line while on the left side of the 
``$t^{}_{\rm GW}=t^{}_{\rm RR,\,v}$'' line keep their orbital orientation during their inspiral
and complete it without abrupt plunge. One can see that, for our choice of
parameters, this is the case for all prograde orbits. On the other hand,
retrograde orbits can cross the ``$t^{}_{\rm GW}=t^{}_{\rm r,\,peri}$'' line while RR
is still effective enough to change their orientation during inspiral. However,
the effect of RR on retrograde orbits cannot reduce significantly the total
EMRI rate and may even increase it slightly because (1) these orbits contribute
less that the prograde ones (and more to the plunge rate) and (2)
statistically, RR is more likely to make the orbit become less retrograde which
pushes down the separatrix.

Finally, we also note that for the other proposed mechanism to produce EMRIs,
the tidal separation of a binary containing a compact object
\citep{MillerEtAl05,Amaro-SeoaneEtAl07}, the captured objects typically have
much lower eccentricities and smaller semi-major axis. Therefore, they cross
``$t^{}_{\rm GW}=t^{}_{\rm r,\,peri}$'' line and start their inspiral, with orbital
parameters well above the uppermost separatrix (for retrograde orbits).  As the
GW-driven trajectory in the $e-a$ plane is basically parallel to the
separatrices, there is no danger of a premature plunge, even though RR has ample
time to flip the orbital orientation during inspiral.

\section{Conclusions}
\label{sec.conclusions}

In this article we have addressed the problem of direct plunges and MBHs. If
this is a Schwarzschild MBH, the compact object will plunge through the horizon
and will hence not contribute to the mapping of space and time around the MBH,
contrary to an EMRI, which describes thousands of cycles before it merges with
the central MBH. On the other hand, the event rate of plunges is much larger
than that of EMRIs, as a number of different studies by different authors using
different methods find.

Up to now spin effects of the central MBH have been always ignored. Hence, the
question arises, whether a plunge is really a plunge when the central MBH is
spinning. This consideration has been so far always ignored.

So as to estimate EMRI event rates, one needs to know whether the orbital
configuration of the compact object is stable or not, because this is the
kernel of the difference between an EMRI and a plunge.  In this paper we take
into account the fact that the spin makes the LSO to be much closer to the
horizon in the case of prograde orbits but it pushes it away for retrograde
orbits.  Since the modifications introduced by the spin are not symmetric with
respect to the non-spinning case, and they are more dramatic for prograde
orbits, we prove that the inclusion of spin increases the number of EMRI events
by a significant factor. The exact factor of this enhancement depends on the
spin, but the effect is already quite important for spins around $s \sim 0.7$.

We also prove that these fake plunges, ``our'' EMRIs, do spend enough cycles
inside the band of eLISA to be detectable, i.e. they are to be envisaged as
typical EMRIs.  We note here that whilst it is true that EMRIs very near the
new separatrix shifted by the spin effect will probably contribute not enough
cycles to be detected, it is equally true for the old separatrix
(Schwarzschild, without spin).  In this sense, we find that the spin increases
generically the number of cycles inside the band for prograde EMRIs in such a
way that EMRIs very near to the non-spin separatrix, which contributed few
cycles, become detectable EMRIs.  In summary, spin increases the area, in
configuration space of detectable EMRIs.  We predict thus that EMRIs will be
highly dominated by prograde orbits.

Moreover, because spin allows for stable orbits very near the horizon in the
prograde case, the contribution of each cycle to the SNR is significantly
bigger than each cycle of an EMRI around a non-spinning MBH.

We then show that vectorial resonant relaxation will not be efficient enough to
change prograde orbits into retrogrades once GW evolution dominates (which
would make the EMRIs plunge instantaneously, as they would be in a non-allowed
region of phase space).

These new kind of EMRIs we describe here, high-eccentric EMRIs, are produced by
two-body relaxation and, as such, they are ignorant of the Schwarschild
barrier.  While low-eccentricity EMRIs run into the problem of having to find a
way to travers this barrier, our ``plunge-EMRIs'' do not. We predict that EMRI
rates will be dominated by high-eccentricity binaries, with the proviso that
the central MBH is Kerr.

\section*{Acknowledgments}

It is a pleasure to thank Tal Alexander for hints on the intricacy of resonant
relaxation. PAS and CFS thank Clovis Hopman for an enlightening discussion in
Amsterdam at Dauphine on the mean free path of drunkards walking between two
streams.  We also thank Bernard Schutz, for fruitful discussions. PAS is
indebted with Francine Leeuwin for her extraordinary support during his visits
in Paris. This work has been supported by the Transregio 7 ``Gravitational Wave
Astronomy'' financed by the Deutsche Forschungsgemeinschaft DFG (German
Research Foundation).  CFS acknowledges support from the Ram\'on y Cajal
Program of the Spanish Ministry of Education and Science, from a Marie Curie
International Reintegration Grant (MIRG-CT-2007-205005/PHY) within the 7th
European Community Framework Program, and from contracts AYA-2010-15709 and
FIS2011-30145-C03-03 of the Spanish Ministry of Science and Innovation and
2009-SGR-935 of the Catalan Agency for Research Funding (AGAUR).  We
acknowledge the computational resources provided by CESGA (CESGA-ICTS-$221$).

\end{document}